\documentclass[aps,12pt,showpacs,nofootinbib,amsmath, floatfix]{revtex4}

\usepackage{epsfig}
\usepackage{graphicx}
\usepackage{amsmath}
\usepackage{amsfonts}
\usepackage{amssymb}
\usepackage{color}
\usepackage{amssymb,amsthm,amscd, amsbsy, array}

\def\bc{\begin{center}}
\def\nno{\nonumber}
\def\ec{\end{center}}
\def\be{\begin{eqnarray}}
\def\ee{\end{eqnarray}}

\newcommand{\omits}[1]{}

\definecolor{dyellow}{rgb}{1.,0.8,.0}
\definecolor{myblue}{rgb}{.1,.1,.7}
\definecolor{dcyan}{rgb}{.0,.6,.6}
\definecolor{dmagenta}{rgb}{0.6,0.0,0.6}
\definecolor{brown}{rgb}{0.6,0.2,0.}
\definecolor{darkblue}{rgb}{.0,.0,0.5}
\definecolor{darkred}{rgb}{0.75,0.0,0.0}
\definecolor{orange}{rgb}{1.,.6,.0}
\definecolor{dorange}{rgb}{0.8,.4,.0}
\definecolor{darkgreen}{rgb}{0.0,0.6,0.0}
\definecolor{purple}{rgb}{.4,.0,.4}

\def\R{I\!\!R}

\def\N{N \hspace{-0.7em}_{_{\sim}} ~}

\def\eps{\epsilon}
\def\ka{\kappa}

\def\d#1#2{\frac{\displaystyle #1}{\displaystyle #2}}
\def\r{\partial}
\def\P{{\bf P}}
\def\K{{\bf K}}
\def\H{{H}}
\def\J{{\bf J}}
\def\N{{\bf N}}
\def\M{{\it M}}
\def\R{{\it R}}

\newcommand{\UWF}{Umov-Weyl-Fock}

\newcommand{\dS}{$d{S}$}
\newcommand{\AdS}{${A}d{S}$}
\newcommand{\Mink}{$Mink$}

\newcommand{\PoRcr}{${P}o{R}_{c,l}$}
\newcommand{\PoR}{principle of relativity}
\newcommand{\PoRcl}{principle of relativity with $(c,l)$}

\newcommand{\SR}{special relativity}

\newcommand{\LFT}{${LFT}$}

\newcommand\btd{\raise 2pt
\hbox{$\hat\bigtriangledown$}\hskip 1.5pt}
\newcommand\bt{\raise 2pt
\hbox{$\bigtriangledown$}\hskip 1.5pt}

\def\PRD{{\it Phys. Rev.}~{\bf D}}

\def\PLA{{\it Phys. Lett.}~{\bf A}}
\def\PLB{{\it Phys. Lett.}~{\bf B}}
\def\GRG{{\it Gen. Rel. Grav. }}

\def\CTP{{\it Comm. Theor. Phys. }}

\begin{document}

\title{The Principle of Relativity, Kinematics and Algebraic Relations}

\author{{Han-Ying Guo}$^{1}$}
\email{hyguo@itp.ac.cn}
\author{{Chao-Guang Huang}$^{2,3}$}
\email{huangcg@mail.ihep.ac.cn}
\author{{Hong-Tu Wu}$^{4}$}
\email{htwu97@gmail.com}
\author{{Bin Zhou}$^{5}$} \email{zhoub@bnu.edu.cn}

\affiliation{%
${}^1$ Institute of Theoretical Physics, Chinese Academy of
Sciences, Beijing 100190, China}%
 \affiliation{%
 ${}^2$ Institute of
High Energy Physics, Chinese Academy of Sciences, Beijing 100049,
China}
\affiliation{%
${}^3$ Theoretical Physics Center for Science Facilities, Chinese
Academy of Sciences, Beijing 100049, China}
\affiliation{%
${}^4$Department of Mathematics, Capital Normal University, Beijing
100048, China}
\affiliation{%
${}^5$Department of Physics, Beijing Normal University, Beijing
100875, China.}
\date{December 12, 2008}

\begin{abstract}
Based on the principle of relativity and the postulate on
universal invariant constants ($c,l$), all
 possible kinematics
  can  be set up with
sub-symmetries of the Umov-Weyl-Fock transformations for the
inertial motions. Further, in the combinatory approach, all these
 symmetries are intrinsically related to each other, e.g. to
 the very important \dS\ kinematics for the cosmic scale physics. %
\end{abstract}

\pacs{%
03.30.+p, 
11.30.Cp  
02.20.Sv  
02.40.Dr, 
}

\maketitle

\tableofcontents

\section{Introduction}

In order to face the challenges of  precise cosmology \cite{SN,
WMAP},  it is needed to re-examine the principles of Einstein's
theory of relativity from the very beginning \cite{srt}. Especially,
whether the \PoR\ should be generalized to the de Sitter (\dS)
spacetime, to which our universe is possibly accelerated expanding
asymptotically.

This is definitely the case and there are two other
kinds of \SR\ in the \dS/anti-\dS\ (\AdS) spacetimes of 
radius $l$ \cite{Lu,LZG, Hua, psr, BdS, TdS,IWR,NH,Lu05, C3,Yan,
OoI, Lu80, duality} based on the \PoR\ and the postulate on
universal constants: {\it There are two universal invariant
constants: the speed of light $c$ and  the length $l$}.  If  the
radius $l$ is linked to the cosmological constant $\Lambda$ from
precise cosmology, $l^2\cong 3/\Lambda$,  the \dS\ \SR\ provides the
new kinematics for the cosmic scale physics \cite{BdS, OoI, Lu80,
duality}. Hereafter, the principle with the
postulate is simply called the \PoRcl\ (the \PoRcr\ for short). Moreover, since the  inertial motion equation 
 is invariant under time reversal and  space inversion,
the \PoRcr\ connotatively require the invariance under them.

 Since the physical space and time coordinates must have
right dimensions,
 the \PoRcr\ should always make sense.
  Although $l$ does not explicitly appear in Einstein's \SR\ and so do
  $l$ and $c$ in Newtonian mechanics, these parameters should  be there implicitly.
Actually, all possible kinematics are deeply related to each other
based on the \PoRcr.

This has been shown for the Poincar\'e algebra $\frak{iso}(1,3)$ in
Einstein's \SR\ in the combinatory approach \cite{srt}. It is
started with the most general transformations among inertial
motions, i.e. the linear fractional transformations with common
denominators (\LFT s) of twenty-four parameters, called the
Umov-Weyl-Fock-transformations $\cal T$, which form the inertial
motion transformation group ${IM}(1,3)$ with inertial motion algebra
$\mathfrak{im}(1,3)$. As its  subalgebra,  the Poincar\'e algebra
contains the Lorentz algebra $\frak{so}(1,3)$ for spacetime
isotropy. Then $IM(1,3)$  is denoted by $IM_L(1,3)$. For
 the \dS/\AdS\ \SR\ based on the \PoRcr,
there are the \LFT s with Lorentz isotropy of \dS/\AdS-group
$SO(1,4)/SO(2,3)$ among the inertial motions in the Beltrami model
of \dS/\AdS-spacetime, respectively\cite{BdS}. So, these \LFT s also
form subgroups of ${IM}_L(1,3)$. However, there are thirty
parameters for Poincar\'e, \dS\, and \AdS\ groups in total. This
seems a puzzle if  no relations among them. Actually, three kinds of
special relativity form a \SR\ triple with the common Lorentz
isotropy \cite{srt}. Further, the translations for transitivity in
$\mathfrak{so}(1,4),\mathfrak{so}(2,3),\mathfrak{iso}(1,3)$, denoted
as $\frak{d}_+,\frak{d}_-,\frak{p}$, are related  in the space of
$\frak{im}_L(1,3)$: The  Euclid translations $\P_\mu \in \frak{p} $
can be given by the plus combination  of
 the Beltrami translations $\P^\pm_i\in \frak{d}_\pm$, i.e.
 $\P_\mu=\frac 1 2(\P^+_\mu+ \P^-_\mu)$. While the
  other generators  transform them from one to another. Moreover, corresponding groups and spacetimes
  are also related. 
  Further, the minus combination leads
to the pseudo-translations $\P'_\mu=\frac 1 2 (\P^+_\mu-\P^-_\mu)$
for the second Poincar\'e algebra
 $\frak{p}_2$. Thus, in the
 space of $\mathfrak{im}(1,3)$, the  \dS/\AdS\
algebras with  an algebraic doublet $(\frak{d}_+,\frak{d}_-)$ can
be transformed to  two Poincar\'e
 algebras with a
doublet $(\frak{p},\frak{p}_2)$ and vice versa, while both $c$ and
$l$ are unchanged. This shows that the combinatory approach is
very different from the contraction approach \cite{IW,Bacry}.

In this paper, by means of the combinatory approach we show that if
the Lorentz isotropy is relaxed to the space isotropy of  rotation
algebra $\frak{so}(3)$, all other
 kinematic algebras of ten generators \cite{Bacry} can also be set up based on the \PoRcr,
 in addition to the
relativistic algebras and their geometric counterparts, i.e. three
algebras $\mathfrak{so}(5),\mathfrak{so}(4,1), \mathfrak{iso}(4)$
denoted as $\frak{r}, \frak{l}, \frak{e}$ for
 4-Riemann/Lobachevsky/Euclid geometry, respectively. And all these
 symmetries are intrinsically related to each other, e.g. to
 the \dS\ algebra that is very important for the cosmic scale physics.

As usual, in addition to the space isotropy algebra
$\mathfrak{so}(3)$ of generators $\J_i$ without dimension, there are
four  types of time and space translations: the Beltrami ones and
the Euclid ones  with their pseudo counterparts: $\{ {\cal H}\}:=\{
H^\pm, \H, \H'\}$ of dimension $[\nu], \nu:=c/l$ called the
Newton-Hooke  constant, and $\{ \P\}:=\{\P^\pm_i, \P_i, \P_i'\}$ of
dimension $[l^{-1}]$, as scalars or vectors  of $\mathfrak{so}(3)$,
respectively, as well as four types of boosts: $\{ \K\}:=\{\K_i,
\N_i,
 \K^{\mathfrak{g}}_i,\K^{\mathfrak{c}}_i\}$ of
the Lorentz,
 geometric,  Galilei and  Carroll boost
of dimension $[c^{-1}]$ as  vector representations of
$\mathfrak{so}(3)$.
Namely\footnote{Hereafter, e.g. $[\J,\P]=\P$ is a shorthand 
of $[\J_i,\P^\pm_j]=-\epsilon_{ij}^{~~k}\P^\pm_k$ etc. And
$\epsilon_{123}=-\eps_{12}^{\ \ 3}=1, \ \ \eta_{ij}=-\delta_{ij},\,\
i,j=1,2,3$.},
\be\label{eq: JHPK}%
[\J,\J]=\J, \quad[\J, {\cal H}]=0, \quad [\J, {\P}]={\P}, \quad[\J, {\K}]={\K}.%
\ee%

In fact, for the \SR\ triple, there is an algebraic quadruplet
$(\frak{d}_+, \frak{d}_-, \frak{p},\frak{p}_2)$ made of four
triplets and six doublets simplicially. If we replace the Lorentz
boost $\K_i$ by the geometric one $\N_i$, the isotropy algebra
$\frak{so}(1,3)$ becomes $\frak{so}(4)$. With suitable translations,
it follows three algebras $\frak{r},\frak{l},\frak{e}$ of
Riemann/Lobachevsky/Euclid geometry, respectively, and the
corresponding algebraic multiplets. If we require both Euclid
translations $H$ and $\P_i$, it follows the Galilei and the Carroll
algebra $\mathfrak{g}, \mathfrak{c}$ of a doublet $(\frak{g},
\frak{c})$ with the Galilei boost $\K^{\mathfrak{g}}_i$ and the
Carroll boost $\K^{\mathfrak{c}}_i$, respectively. If we replace $H$
by the Beltrami one $H^+, H^-$ and still keep $\P_i$, the
Newton-Hooke/anti-Newton-Hooke ($NH_\pm$) algebras
$\mathfrak{n}_\pm$ follow with the Galilei boost
$\K^{\mathfrak{g}}_i$, respectively. While, if we replace $\P_i$ by
the Beltrami  ones $\P^+_i,\P^-_i$ and still keep $H$, the
Hooke-Newton/anti-Hooke-Newton ($HN_\pm$) algebras
$\mathfrak{h}_\pm$ follow with the Carroll boost
$\K^\mathfrak{c}_i$, respectively \footnote{In \cite{Bacry} and in
literatures, both are called the para-Poincar\'e algebras.}.
Further, other algebras can also be given easily by the combinatory
approach in the
 space of $\mathfrak{im}(1,3)$. As  the
second Poincar\'e algebra $\frak{p}_2$ and the doublet
$(\frak{p},\frak{p}_2)$, there are always the second algebra for
each algebra with the pseudo-translations $H'$ or $\P'_i$ and a
doublet for each pair of them, except  the \dS/\AdS\ algebras
$\frak{d}_\pm$ and their geometric counterparts $\frak{r},\frak{l}$.
Moreover, there are also algebraic multiplets made of lower
multiplets simplicially with respect to the $\mathfrak{im}(1,3)$, so
all these kinematic algebras are intrinsically related. For example,
by combining the boosts of the Galilei and Carroll algebras
$\frak{g,c}$, it follows the Poincar\'e algebra $\frak{p}$, which
leads to the \dS\ algebra $\frak{d}_+$ directly by the combination
of the translations and pseudo-translations of $\frak{p}_2$.

This paper is arranged as follows. In section \ref{sec: UWF}, we
briefly recall the \UWF\ transformations for the \PoRcr\
 and how to
 get the algebras of three kinds of \SR, their triple with algebraic relations  and
 their geometric counterparts.
 In  section \ref{sec: NRA},
we first get the Galilei and Carroll algebras $\mathfrak{g},
\mathfrak{c}$ as well as their second ones. Then we  get the
$NH_\pm$ algebras $\mathfrak{n}_\pm$, the $HN_\pm$ algebras
$\mathfrak{h}_\pm$ and other non-relativistic algebras as well as
their second ones by means of the combinatory approach. In section
\ref{sec: NRT}, we illustrate  the  relations among these kinematics
and some physical implications. Finally, we end with some remarks.


\section{The  Inertial Motion Algebra, the
Special Relativity triple and Its Geometric Counterpart}
\label{sec: UWF}

\subsection{The principle of relativity with $(c,l)$ and the \UWF-transformations}

In the inertial coordinate frames ${\cal F}:=\{S(x)\}$, a free
particle takes inertial motion{\footnote{ Conventionally, the
concepts are depended on certain metric introduced. These had been
generalized in \cite{srt, Umov, Weyl, Fock}.}}
 \be\label{eq:uvm}%
 x^i=x_0^i+v^i(t-t_0),~~
v^i=\frac{dx^i}{dt}={\rm consts}.~~ i=1, 2, 3. %
\ee%
What are the most general transformations ${\cal T}:=\{T\}$
\be\label{eq:FL}%
{\cal T}\ni T:\quad {x'}^\mu=f^\mu( x, T), ~ x^0=ct, ~ \mu=0,\cdots, 3,%
\ee%
that keep the inertial motion (\ref{eq:uvm}) invariant?

It can be proved \cite{Umov,Weyl,Fock,duality}: {\it The most
general form of the transformations (\ref{eq:FL}) are  the \LFT s
\begin{equation}
{ T}:\quad  l^{-1}x'^\mu = \frac{A^\mu_{\ \nu} l^{-1} x^\nu +
b^\mu}{c_\lambda  l^{-1}x^\lambda + d} \label{eq:LFT}
\end{equation}
and
\begin{equation}
  det\ { T}=\left| \begin{array}{rrcrr}
    A    & b^t \\
    c
     & d
  \end{array} \right|
  = 1,
\label{eq:det}
\end{equation}
 where $A=\{A^\mu _{~\nu}\}$  a $4\times 4$ matrix, $b,c$  $1\times 4$
 matrixes, $d\in R$
and superscript
  $\small{^ t}$
 for the transpose.}

 Clearly,  all Umov-Weyl-Fock
transformations $T\in {\cal T}$  form the inertial motion
transformation group $IM(1,3)$ of twenty-four generators with its
algebra $\mathfrak{im}(1,3)$. Actually, the inertial motion
(\ref{eq:uvm}) may be viewed as a straight line and the most general
transformations among straight lines form the real projective group
${RP}(4)$ (see, e.g. \cite{Hua, psr}). But, for keeping the
orientation the antipodal identification should not be taken
\cite{BdS}.

 Further, all
algebraic relations must keep the invariance of time reversal and
space inversion.

\subsection{Three kinds of special relativity as a triple and  algebraic relations }

As in \cite{srt}, we  first consider the Lorentz isotropy   and the
transitivity  by the Euclid translations of Abelian group $R(1,3)$
generated by  $\H$ and $\P_i$ for
  coordinates afterwards. Then, it follows the Poincar\'e transformations
of
 $ISO(1,3)=R(1,3)\rtimes SO(1,3)$  and
 both isotropy and transitivity  make the Minkowski (\Mink) spacetime as a
4d homogeneous space $M=ISO(1,3)/SO(1,3)$.
 The Poincar\'e algebra  $\mathfrak{iso}(1,3)$, i.e. $\mathfrak{p}$, is generated by
 the set
$\{T\}^{\mathfrak{p}}:=(H,\P_i, \K_i,\J_i)$
\be\nno
 H=\partial_t, ~~ {\mathbf P}_i = \partial_i,&&
 \K_i=t \partial_i -c^{-2} x_i
 \partial_t, \\\label{eq: Pgenerator}
 \J_i=\d 1 2
 \epsilon_{i}^{~~jk}L_{jk},&&L_{jk}:=x_j\partial_k-x_k\partial_j,
  \ee%
where $x_\mu:=\eta_{\mu\nu}x^\nu, \eta_{\mu\nu}={\rm
diag}(1,-1,-1,-1)$  and $\epsilon_i^{\ jk} = \epsilon_{ijk}$.

In order to get the \dS/\AdS\  transformations with Lorentz
isotropy,  the  Euclid translations $(H, \P_i)$ should be replaced
by the Beltrami translations $(H^\pm, \P^\pm_i)$ \cite{srt},  respectively%
\be\label{eq: Ppm}%
 H^\pm=\partial_t\mp \nu^{2}t x^\nu\partial_\nu, \quad
 {\mathbf P}^\pm_i =\partial_i\mp l^{-2}x_i
 x^\nu\partial_\nu.
\ee%
 Thus, from the Beltrami model of
\dS/\AdS-spacetime \cite{srt}, it follows  the \LFT s, denoted by
${\cal S}_\pm$, of \dS/\AdS-group $SO(1,4)/SO(2,3) \subset
{IM}_L(1,3)$, which keep the coordinate domain conditions and the
Beltrami metrics invariant, respectively. Actually,  the
Beltrami-\dS/\AdS-spacetime ${\cal B}_\pm$ is also a homogeneous
space ${\cal B}_\pm \cong {\cal S}_\pm/SO(1,3)$, respectively. Then
the sets $\{T\}^{\mathfrak{d}_\pm}:=(H^\pm,\P^\pm_i, \K_i,\J_i)$
span the \dS/\AdS\ algebra $\mathfrak{so}(1,4),\mathfrak{so}(2,3)$
or $\mathfrak{d}_\pm $, respectively. And the \dS/\AdS\ 
doublet $(\frak{d}_+,\frak{d}_-)$ follows.

Further, in the space of  $\mathfrak{im}_L(1,3)$, the 
translations are related by%
\be\label{eq: P}%
   H=\d 1 2 (\H^++\H^-),\quad {\P}_i :=\d 1 2 \left({\P}^+_i+{\mathbf P}^-_i
   \right),
 \ee%
where the constant $l$  is hidden,  while the dimensions of the
generators and algebra are still kept.

In addition to all these generators of
$\frak{d}_+,\frak{d}_-,\frak{p}$ for three kinds of \SR,
 there are still  other ten generators in the
 $\mathfrak{im}_L(1,3)$ as follows %
 \begin{equation}
\begin{split}
\N_i = t \partial_i + c^{-2} x_i
\partial_{t}, \quad
R_{ij} =R_{ji}= x_i \partial_j + x_j \partial_i, (i< j),\quad%
M_\mu = x^{(\mu)} \r_{(\mu)},
\end{split}
\end{equation}
where no summation is taken for  repeated  indexes in brackets.

 Together with the \dS/\AdS\  relations, the rest
 non-vanishing  relations of
 $\mathfrak{im}_L(1,3)$ of the set
 $\{T\}^\mathfrak{im}:=(\H^\pm, \P^\pm_i,\J_i,\K_i,\N_i,\M_0,\M_i,\R_{ij})$
 read:
 \omits{
\begin{equation}
\begin{split}
[\H^+, \H^-] =  2{ \nu^2}& \left(\M_0 + \Sigma_\ka \M_\ka\right),\\
[\P^+_i, \P^-_j] = (1-\delta_{(i)(j)})l^{-2} & R_{(i)(j)} - 2l^{-2}
\delta_{i(j)} \left(\M_{(j)}+\Sigma_\ka \M_\ka\right), \\
[\K_i, \N_j] =(\delta_{(i)(j)}-1) & c^{-2} R_{(i)(j)} -
2\delta_{(i)j}c^{-2}\left(M_{0} - M_{(i)}\right),\\
[\H^\pm, \M_0]= \H^\mp,\quad & [\P^\pm_i, \M_j]= \delta_{i(j)}\P^\mp_{(j)},\\
[\P^{\pm}_i, \R_{jk}]
=-\delta_{ij}\P^{\mp}_k-\delta_{ik}\P^{\mp}_j,\quad &
[L_{ij}, \M_k]=\delta_{j(k)}  R_{i(k)}-\delta_{i(k)} R_{j(k)}, \\
[L_{ij},\R_{kl}]=
2(\delta_{ik}\delta_{jl}+\delta_{il}\delta_{jk})\left(\M_i - \M_j
\right)& + \delta_{ik}\R_{jl} +
\delta_{il}\R_{jk}-\delta_{jk} R_{il}-\delta_{jl} R_{ik},\\
[\K_i, \M_0]=-\N_i, \quad & [\K_i, \M_j]= \delta_{i(j)} \N_{(j)},\\
[\K_i, \R_{jk}]= -\delta_{ij} \N_k-\delta_{ik}\N_j, \quad &[\N_i,
\M_0]= -\K_i, \\
[\N_i, \M_j]= \delta_{i(j)}\K_{(j)}, \quad & [\N_i
,\R_{jk}]=-\delta_{ij} \K_k-\delta_{ik} \K_j, \\
 [R_{ij}, \M_k]=\delta_{i(k)}&L_{j(k)} + \delta_{j(k)}L_{i(k)},\\
[\R_{ij},R_{kl}] =
-\delta_{ik}R_{jl}&-\delta_{il}R_{jk}-\delta_{jk}R_{il}-\delta_{jl}R_{ik}.\end{split}\label{im4}
\end{equation}}
\begin{eqnarray}
\begin{array}{l}
\, {[}\P^+_i, \P^-_j{]} =(1-\delta_{(i)(j)})l^{-2} R_{(i)(j)} -
2l^{-2}
\delta_{i(j)} (\M_{(j)}+\Sigma_\ka M_\ka), \\
\begin{array}{ll}
{[}\P^\pm_i, \M_j{]}= \delta_{i(j)}\P^\mp_{(j)},&
[\P^{\pm}_i, \R_{jk}]=-\delta_{ij}\P^{\mp}_k-\delta_{ik}\P^{\mp}_j,\\
{[}\H^+, \H^-{]} =  2{ \nu^2} \left(\M_0 + \Sigma_\ka \M_\ka\right),
&
[\H^\pm, \M_0]= \H^\mp,  \\
{[}\K_i, \M_0{]}=-\N_i, &   [\K_i, \M_j]= \delta_{i(j)} \N_{(j)},\\
{[}\K_i, \R_{jk}{]}= -\delta_{ij} \N_k-\delta_{ik}\N_j,& [\N_i, \M_0]= -\K_i,\\
{[}\N_i, \M_j{]}= \delta_{i(j)}\K_{(j)},&
{[}\N_i,\R_{jk}{]}=-\delta_{ij} \K_k-\delta_{ik} \K_j, \\
{[}\K_i,\, \N_j{]} =(\delta_{(i)(j)}-1)  c^{-2} R_{(i)(j)} -&
2\delta_{(i)j}c^{-2}\left(M_{0} - M_{(i)}\right),\\
{[}L_{ij}, \M_k{]}=\delta_{j(k)}  R_{i(k)}-\delta_{i(k)}R_{j(k)} ,&
{[}R_{ij}, \M_k{]}=\delta_{i(k)}L_{j(k)} + \delta_{j(k)}L_{i(k)},
\end{array}\\
\,
{[}L_{ij},\R_{kl}{]}=2(\delta_{ik}\delta_{jl}+\delta_{il}\delta_{jk})\,
(\M_i  - \M_j
) + \delta_{ik}\R_{jl} +\delta_{il}\R_{jk}-\delta_{jk} R_{il}- \delta_{jl}R_{ik}, \\
\, {[}\R_{ij},R_{kl}{]} = -
\delta_{ik}R_{jl}-\delta_{il}R_{jk}-\delta_{jk}R_{il}-\delta_{jl}R_{ik}.
\end{array} \label{im4}
\end{eqnarray}%
The algebra $\mathfrak{im}_L(1,3)$ is the closed algebra for the
doublet $(\frak{d}_+,\frak{d}_-)$.

 Clearly, in addition to
 the common Lorentz  isotropy and the relation between the
 Euclid and Beltrami translations $\P_{\mu}, \P^\pm_\mu$, four generators $(M_0,M_i)$
 and six generators $(\N_i, R_{ij})$  of
   $\mathfrak{im}_L(1,3)$
 exchange these  translations
  from one to another. Thus,
three kinds of \SR\ act as a whole in $IM_L(1,3)$, called {\it the
\SR\ triple}\cite{srt}.

There are rich
 algebraic
relations in the triple.
 From  Eq. (\ref{eq: P}), it
follows naturally
the pseudo-time/space-translations by linear combination %
\be\label{eq: H'P'}%
 H'=\d 1 2(H^+-H^-)=-\nu^2 t 
 x^\mu\partial_\mu, \quad
 {\P}'_i :=\d 1 2 \left({\P}^+_i-{\mathbf P}^-_i \right)=-
l^{-2} x_i 
x^\nu \r_\nu.
 \ee%
 The set  $\{T\}^{\mathfrak{p}_2}:=(H',\P'_i, \K_i,\J_i)$
 spans  the second Poincar\'e algebra $\mathfrak{p}_2$. It is important that this algebra preserves the \Mink-light cone at the
 origin.

 Thus, dual
 to the \dS/\AdS\ doublet $(\mathfrak{d}_+,\frak{d}_-)$ there is the Poincar\'e doublet
 $(\mathfrak{p},\mathfrak{p}_2)$ and  both can be transformed from one to another by either the
linear combinations of the generators or the algebraic relation in
$\mathfrak{im}(1,3)$. In fact,   with respect to the
$\mathfrak{im}(1,3)$, there is a
  \dS/\AdS/Poincar\'e algebraic quadruplet
  $(\mathfrak{d}_+, \mathfrak{d}_-, \mathfrak{p},\mathfrak{p}_2)$ made simplicially of 4 algebraic triplets $(\mathfrak{d}_+, \mathfrak{d}_-, \mathfrak{p})$
  , $(\mathfrak{d}_+, \mathfrak{d}_-, \mathfrak{p}_2)$, $(\mathfrak{d}_+,  \mathfrak{p},\mathfrak{p}_2)$
  and $(\mathfrak{d}_-, \mathfrak{p},\mathfrak{p}_2)$, as well as
 6 doublets $(\mathfrak{d}_+, \mathfrak{d}_-),
  (\mathfrak{p},\mathfrak{p}_2)$,
  $(\mathfrak{d}_+, \mathfrak{p}), (\mathfrak{d}_+,\mathfrak{p}_2)$,
  $(\mathfrak{d}_-, \mathfrak{p})$ and $(\mathfrak{d}_-,\mathfrak{p}_2)$.
And  the closed algebra of the quadruplet is also the
  $\mathfrak{im}(1,3)$.

\subsection{The geometry triple with $\frak{so}(4)$ isotropy and algebraic relations}
If we replace the Lorentz boost $\K_i$  by the geometric boost
$\N_i$ and at the same time,  due to  the signature $-2$ of
$\eta_{\mu\nu}$ we  adjust the Beltrami time and space translations
suitably,  the relativistic algebras turn to their geometric
counterparts. That is, the Riemann/Lobachevsky/Euclid algebras
$\mathfrak{so}(5),\mathfrak{so}(4,1),\mathfrak{iso}(4)$  or
$\mathfrak{r},\mathfrak{l},\mathfrak{e}$  with generator sets
$\{T\}^{\mathfrak{r}}:=(H^{-}, \P^+_i, \N_i, \J_i)$,
$\{T\}^{\mathfrak{l}}:=(H^+, \P^-_i, \N_i, \J_i)$ and
$\{T\}^{\mathfrak{e}}:=(H, \P_i, \N_i, \J_i)$  for 4d Riemann
sphere, Lobachevsky hyperboloid and
 Euclid space, respectively. Thus, there is a geometry triple with rich
 algebraic relations.

First, there is also the second Euclid algebra $\mathfrak{e}_2$ with
generator set $\{T\}^{\mathfrak{e}_2}:=(H', \P'_i, \N_i, \J_i)$.
Then, corresponding to  the relativistic  quadruplet, there is a
geometric algebraic quadruplet $(\mathfrak{r}, \frak{l},
\mathfrak{e},\mathfrak{e}_2)$, too.

It should be noticed that  the geometry triple is  in the sense of
 the straight-lines as the geometric `inertial motions' in the inertial
  frames with one geometric
`time' coordinate $t$ and three space coordinates $x^i$. Although
two universal constants $(c,l)$ are still there, the signatures of
the metrics are intrinsically changed to the geometric one,
respectively.

\begin{table}[thp]
\caption{\quad All possible relativistic, geometric and
non-relativistic kinematic algebras} {\scriptsize
\begin{tabular}{|c c c c c c c c|}
\hline Algebra & Symbol & Generator set\footnote{All  generators
and commutators have right dimensions expressed by the universal
constants $c, l$ or $\nu$.}
  & $[\cal H,\P]$ & $[\cal H,\K]$ & $[\P,\P]$ &$[\K,\K]$ &$[\P,\K]$\\
\hline \dS  &$\mathfrak{d}_+$ & $(H^+, \P^+, \K, \J)$ & $\nu^2\K$ &
$ \P$ & $l^{-2}\J$ &
$-c^{-2}\J$  &$c^{-2}\cal H$\\
\AdS & $\mathfrak{d}_-$ &  $(H^-, \P^-, \K, \J)$ & $-\nu^2\K $ & $
\P$ & $-l^{-2}\J $
& $-c^{-2}\J$ & $c^{-2}\cal H$\\
{\it Poincar\'e} &$\begin{array}{c}\mathfrak{p}\\
\mathfrak{p}_2\end{array}$ & $\begin{array}{c}(H, \P, \K, \J)\\
\{H', \P', \K, \J\}\end{array}$ & 0 & $\P$ & 0 &$-c^{-2}\J$&$c^{-2}\cal H$\\
\hline {\it Riemann} & $\mathfrak{r}$&$(H^-, \P^+, \N,\J)$&
$-\nu^2\K$ & \P &$l^{-2}\J$ &$
c^{-2}\J$ &$-c^{-2}\cal H$\\
{\it Lobachevsky}&$\mathfrak{l}$&$(H^+, \P^-, \N,\J)$& $\nu^2\K$ &
\P & $-l^{-2}\J$
 &$c^{-2}\J$ &$-c^{-2}\cal H$\\
{\it Euclid}&$\begin{array}{c}\mathfrak{e}\\
\mathfrak{e}_2\end{array}$&$\begin{array}{c}(H, \P, \N, \J)\\
(H', \P', \N, \J)\end{array}$&0 &\P &0 &$c^{-2}\J$ &$-c^{-2}\cal H$\\
\hline
{\it Galilei}&$\begin{array}{c}\mathfrak{g}\\
\mathfrak{g}_2\end{array}$&$\begin{array}{c}(H, \P, \K^{\frak g},\J)\\
(H', \P', \K^{\frak c},\J)\end{array}$&0 & \P  &  0 &   0 &   0   \\
{\it Carroll}&$\begin{array}{c}\mathfrak{c}\\
\mathfrak{c}_2\end{array}$ & $\begin{array}{c}(H, \P, \K^{\frak c},\J )\\
(H', \P', \K^{\frak g},\J )\end{array}$ & 0& 0& 0& 0&$c^{-2}\cal H$\\
${NH}_+$  & $\begin{array}{c}\mathfrak{n_+}\\
\mathfrak{n}_{+2}\end{array}$ & $\begin{array}{c}(H^+, \P , \K^{\frak g},\J )\\
(H^+, \P', \K^{\frak c},\J )\end{array}$&$\nu^2\K$ &\P & 0 & 0 &0 \\
${NH}_-$  &$\begin{array}{c}\mathfrak{n}_-\\
\mathfrak{n}_{-2}\end{array}$ & $\begin{array}{c}(H^-, \P, \K^{\frak g},\J )\\
(-H^-, \P', \K^{\frak c},\J )\end{array}$ & $-\nu^2\K$ &\P &0 &0 &0 \\
{\it para-Galilei}&$\begin{array}{c}\mathfrak{g}'\\
\mathfrak{g}'_2\end{array}$ & $\begin{array}{c}(H', \P, \K^{\frak g}, \J)\\
(H, \P', \K^{\frak c},\J )\end{array}$ & $\nu^2\K$ & 0 & 0 & 0 &0 \\
$HN_+$&$\begin{array}{c}\mathfrak{h}_+\\
\mathfrak{h}_{+2}\end{array}$&$\begin{array}{c}(H, \P^+, \K^{\frak c},\J )\\
(H', \P^+, \K^{\frak g},\J )\end{array}$&$\nu^2\K$ &0 &$l^{-2}\J$&0 &$c^{-2}\cal H$\\
$HN_-$&$\begin{array}{c}\mathfrak{h_-}\\
\mathfrak{h}_{-2}\end{array}$&$\begin{array}{c}(H, \P^-, \K^{\frak c},\J )\\
(-H', \P^-, \K^{\frak g},\J )\end{array}$&$-\nu^2\K$ &0 &$-l^{-2}\J$&0 &$c^{-2}\cal H$\\
\hline {\it Static}&$\begin{array}{c}\mathfrak{s}\\
\mathfrak{s}_2\end{array}$&$\begin{array}{c}(H^{\frak s}, \P',
\K^{\frak c},\J )\footnote{The generator $H^{\frak s}$ is
meaningful only when the central extension is considered.}\\
(H^{\frak s}, \P , \K^{\frak g},\J )\end{array}$&0&0&0&0&0\\
\hline
\end{tabular}
}

\end{table}


\section{The  Non-relativistic Kinematics and  Algebraic Relations}
\label{sec: NRA}

\subsection{The Euclid time and space translation algebras: Galilei, Carroll
 and counterparts}

The Galilei transformations as a subset of  \UWF\ transformations
(\ref{eq:FL}) can be easily reached from the space isotropy
generated by $\J_i$ with the Galilei boost $\K^{\mathfrak{g}}_i$ at
origin as well as by the time
 and
space translations  $H, \P_i$ for transitivity. The generators can
be easily obtained
\begin{equation}\label{eq: n1}
{\H}
:=\partial_t=\d 1 2 \left(\H^+ + \H^-\right),~~%
{\bf P}_i
:=\partial_i=\d 1 2 \left(\P^+_i + \P^-_i\right),~~%
 {\bf K}_i^{\mathfrak{g}}: = t\partial_i=\d 1 2 \left(\K_i + \N_i\right). %
\end{equation}%
Then the set  $\{T\}^{\mathfrak{g}}:=(\H, \P_i,\K^\mathfrak{g}_i,
\J_i)$ spans  the Galilei algebra $\mathfrak{g}$.

From the viewpoint of combinatory,  dual to the Galilei  boost $
\K_i^{\mathfrak{g}}$, there is the Carroll boost
\be\label{eq: Kcarroll}%
\K_i^{\mathfrak{c}}:=\d 1 2 \left(\K_i - \N_i\right)=-c^{-2} x_i\r_t. %
 \ee%
 Actually, the  set  $\{T\}^{\mathfrak{c}}:=(\H, \P_i,\K^\mathfrak{c}_i, \J_i)$
 spans   the Carroll algebra $\mathfrak{c}\subset\mathfrak{im}(1,3)$. Thus, the
 Carroll kinematics is also based on the \PoRcr. Clearly, there is a Galilei-Carroll doublet
$(\frak{g},\frak{c})$. And, the other two  sets
$\{T\}^{\mathfrak{g}_2}:=(\H',\P'_i, \K^\mathfrak{c}_i, \J_i)$ and
$\{T\}^{\mathfrak{c}_2}:=(\H',\P'_i, \K^\mathfrak{g}_i, \J_i)$ with
pseudo-translations span  the second Galilei and Carroll algebras
$\mathfrak{g}_2$ and $\mathfrak{c}_2$, respectively.

Thus, there are three more Galilei-Carroll doublets
$(\frak{c},\frak{g}_2)$, $(\frak{g},\frak{c}_2)$ and
$(\frak{c}_2,\frak{g}_2)$ as well as the Galilei or Carroll doublets
$(\frak{g},\frak{g}_2)$ or $(\frak{c},\frak{c}_2)$, respectively. In
fact, there is the Galilei-Carroll algebraic quadruplet
$(\frak{g},\frak{g}_2,\frak{c},\frak{c}_2)$ made of 4 triplets and 6
doublets simplicially with respect to the $\frak{im}(1,3)$.

\subsection{The Beltrami time translation algebras:
Newton-Hooke/anti-Newton-Hooke/para-Galilei  and counterparts}
 If we replace the time translation by Beltrami time
translations $\H^\pm$  in (\ref{eq: Ppm}), the  sets
$\{T\}^{\mathfrak{n}_\pm}:=(\H^\pm, \P_j, \K^{\mathfrak{g}}_j,
\J_j)$ span   the $NH_\pm$ algebra $\mathfrak{n}_\pm$, respectively.
And the other two  sets  $\{T\}^{\mathfrak{n}_{\pm 2}}:=(\pm\H^\pm,
\P'_i,
 \K_i^{\mathfrak{c}}, \J_i)$ span  the second
$NH_\pm$ algebras $\mathfrak{n}_{\pm 2}$. Hence, there is the
$NH_\pm$ quadruplet $(\mathfrak{n}_\pm,\mathfrak{n}_{\pm 2})$ made
of four triplets and six doublets simplicially.

Furthermore, again with the pseudo-translation $\H', \P'_i$ in
(\ref{eq: H'P'}) the  sets $\{T\}^{\mathfrak{g}'}:=(H', \P_j,
\K^{\mathfrak{g}}_j, \J_j)$ and $\{T\}^{\mathfrak{g}'_2}:=(\H,
\P'_i, \K^\mathfrak{c}_i, \J_i)$ span  the para-Galilei algebra
$\mathfrak{g}'$ and the second one ${\mathfrak{g}'_2}$,
respectively.
 Then, the para-Galilei doublet
$(\mathfrak{g}',\mathfrak{g}'_2)$ follows.

Thus, the $NH_\pm$-para-Galilei sextuplet
$(\mathfrak{n}_\pm,\mathfrak{n}_{\pm 2},
\mathfrak{g}',\mathfrak{g}'_2 )$ can be made of simplicialy.

\subsection{The Beltrami space translation algebras:
Hooke-Newton/anti-Hooke-Newton/static and counterparts}
\label{sec: NEST}

If we replace the  space translations $\P_i$ by the Beltrami
translations $\P^\pm_i$ in (\ref{eq: Ppm}), still keep the time
translation $\H$,  two  sets $\{T\}^{\mathfrak{h}_\pm}:=(\H, \P^\pm,
\K^{\mathfrak{c}}_i, \J_i)$ span  the $HN_\pm$
 algebra $\mathfrak{h}_\pm\cong \mathfrak{iso}(4)/
 \mathfrak{iso}(1,3)$, respectively. And the $HN_\pm$ doublet
 $(\mathfrak{h}_+,\mathfrak{h}_-)$ follows.

From  algebraic relations of ${\mathfrak{h}_\pm}$, it follows that
the sub-sets $(\P^\pm_i, \J_i)$ form  subalgebras $\mathfrak{so}(4),
 \mathfrak{so}(1,3)$, which keeps a 3d Riemann sphere/Lobachevsky hyperploid $S^3/H^3$,
 respectively, and
 others form 4d Abelian subalgebras. Since the time and  space are also separated
 in analogy to the cases of the $NH_\pm$, it would be
 better to name them $HN_\pm$, rather
 than para-Poincar\'e as was named in \cite{Bacry}. In addition,  exchanging the
 generators $(\H^\pm, \P_i, \K_i^{\frak g})\subset \{T\}^{\mathfrak{n}_\pm}$ in the
$NH_\pm$ algebras by the generators $(\H, \P_i^\pm,
\K_i^\frak{c})\subset \{T\}^{\mathfrak{h}_\pm}$, the $HN_\pm$
algebras follow and vice versa. This is also why the algebras are
called after the name Hooke-Newton,
rather than para-Poincar\'e. %

It can  also be shown that the sets $\{T\}^{\mathfrak{h}_{\pm
2}}:=(\pm\H', \P^\pm_i, \K^\mathfrak{g}_i, \J_i)$ span  the second
$HN_\pm$ algebras $\mathfrak{h}_{\pm 2}\cong \mathfrak{iso}(4)/
 \mathfrak{iso}(1,3)$,
 respectively. Thus, not only more  $HN_\pm$ doublets
 follow, but also the $HN_\pm$ quadruplet
$(\mathfrak{h}_\pm,\mathfrak{h}_{\pm2})$  made of simplicially
follows.

  In the combinatory approach, the so-called static algebra\cite{Bacry} can also be
  reached with%
   \be%
  H^{\mathfrak{s}}=H-H=H'-H'=0%
  \ee%
   by combining the generator sets of $HN_{\pm}$ or $HN_{\pm 2}$ algebras. In other words,
   two  sets
$\{T\}^{\mathfrak{s}}:=(H^{\mathfrak{s}}, \P'_i,
\K^{\mathfrak{c}}_i, \J_i)$ and
$\{T\}^{\mathfrak{s}_2}:=(H^{\mathfrak{s}}, \P_i,
\K^{\mathfrak{g}}_i, \J_i)$ span  two static algebras $\mathfrak{s}$
 and $\mathfrak{s}_2$,
respectively, with a  doublet $(\mathfrak{s},\mathfrak{s}_2)$. In
fact, $\H^{\mathfrak{s}}=0$ is not a generator. If the central
extension is considered, $H^{\mathfrak{s}}$  becomes the one in
\cite{Bacry}.

It is clear that there also exist the $HN_\pm$-static algebraic
sextuplet $({\frak{h}_\pm},{\frak{h}_{\pm 2}}, \mathfrak{s},
{\frak{s}_2})$ made of the triplet $(\mathfrak{h}_+,\mathfrak{h}_-,
\mathfrak{s})$ and so on simplicially.


\section{The  Relations among Kinematics and Some Implications}
\label{sec: NRT}

In  the combinatory approach,  all these kinematical algebras are
related in the  space of $\mathfrak{im}(1,3)$ either by suitably
exchanging some generators or by their algebraic relations with the
generators $(M_0,M_i)$ and $R_{ij}$, no matter whether the algebras
are relativistic, geometric or non-relativistic. Let us illustrate
some of these relations  and relevant physical implications.

As was mentioned, there is a duality between the relativistic and
geometric kinematics and their algebraic relations rather than the
Wick rotation \cite{IWR}. Actually, we may also consider a
relativistic-geometric octuplet with all eight algebras as singlets
simplicially and so on.

There is the Poincar\'e-Euclid-Galilei-Carroll quadruple
$(\mathfrak{p},\mathfrak{e},\mathfrak{g}, \mathfrak{c})$, as well as
its mimic of the second copies,  is a typical
relativistic-geometric-non-relativistic one, in which there is the
Poincar\'e-Euclid doublet $(\mathfrak{p},\mathfrak{e})$ that
implicates the relation between the \Mink-spacetime and the 4d
Euclid space different from the Wick rotation\cite{IWR}. And the
Pincar\'e-Galilei-Carroll triplet $(\mathfrak{p},\mathfrak{g},
\mathfrak{c})$ is a typical relativistic-non-relativistic one. As
for non-relativistic kinematics, the Galilei and  Carroll algebras,
which form a doublet $(\mathfrak{g}, \mathfrak{c})$, indicate that
the Newtonian kinematics is for  the motions  of $v\ll c$,  while
the Carroll kinematics may be for a kind of extremely superluminary
motions with $v \gg c$, having   invariant constant $c$. Since the
algebraic combination of the Galilei boost $\K_i^\frak{g}$ and the
Carroll one $\K_i^\frak{c}$
 leads to the Lorentz boost $\K_i$, it should be for a relativistic one. This is
   just the case
for the Poincar\'e kinematics. Taking into account the
\dS-Poincar\'e triplet $(\frak{d}_+, \frak{p}, \frak{p}_2)$, the
Galilei kinematics can be related to the \dS\ one and vice versa.

Similarly, in the \dS/\AdS-$NH_\pm$ quadruple
$(\mathfrak{d}_+,\mathfrak{d}_-, \mathfrak{n}_\pm)$, there are the
\dS/\AdS-$NH_\pm$  triplets $(\mathfrak{d}_+, \mathfrak{n}_\pm)$ and
$(\mathfrak{d}_-, \mathfrak{n}_\pm)$ together with other two
triplets, which also  make sense, respectively.

As non-relativistic algebras,  the Galilei $\frak{g}$ and $NH_\pm$
$\frak{n}_\pm$ share nine common generators $(\P _j,
\K^{\mathfrak{g}}_j, \J_j)$.
In the Galilei-$NH_\pm$ quadruplet 
$(\mathfrak{g},\mathfrak{g}_2, \mathfrak{n}_\pm)$, the triplet
$(\mathfrak{g}, \mathfrak{n}_\pm)$ made of three doublets
$(\mathfrak{g},\mathfrak{n}_+)$, $(\mathfrak{g},\mathfrak{n}_-)$ and
$(\mathfrak{n}_+,\mathfrak{n}_-)$ and others also indicate certain
sense in physics. In general, for Galilei-$NH_\pm$ there is a
sextuplet $(\mathfrak{g},\mathfrak{g}_2,\mathfrak{n}_\pm,
\mathfrak{n}_{\pm 2})$ made of simplicially by  lower
Galilei-$NH_\pm$ multiplets.

 There are
also the Galilei-para-Galilei-$NH_\pm$ quadruplet
$(\frak{g},{\frak{g}}',\frak{n}_\pm )$ made of lower multiplets
with common generators $({\bf J},{\bf P}, {\bf K})$. Similarly,
their partners
 form another quadruplet $(\frak{g}_2,\ {\frak{g}}_2',\ \frak{n}_{\pm2})$.
Combining them together simplicially, there is  the
Galilei-para-Galilei-$NH_\pm$ octuplet and so on.

For the $HN_\pm$ algebras $({\frak{h}_\pm},{\frak{h}_{\pm 2}})$
 and the static algebras $(\mathfrak{s}, {\frak{s}_2})$, they can also be linked with
 other algebras within certain
multiplets simplicially.

In fact, there are fourteen-type kinematic algebras
\footnote{The static algebra is a special one.} with
twenty-four copies in the space of $\frak{im}(1,3)$.  Thus, there is
a twenty-four-plet for all them made of lower multiplets
simplicially.


\section{Concluding Remarks}

 We have shown that all
possible kinematics can be set up based on the \PoRcr\ and all these
kinematic algebras with ten generators are related.  Thus, their
physical implications should be explored based on the \PoRcr.  Since
the \dS\ kinematics is very important for the cosmic scale physics
characterized by the cosmological constant \cite{BdS, OoI, Lu80},
all other related kinematics may have certain physical meaning in
their own right.

 Different from traditional considerations, two
universal constants $c$ and $l$ always exist there, which make sense
at least for right dimensions of coordinates, generators and
algebraic relations.

Those algebras  with the pseudo-translations $H'$ and $\P'_i$ cannot
make transitivity in time and space and  might be ignored in
space-time physics. However, not only they do make sense in
combinatory, but also still play some physical roles. For example,
the second Poincar\'e algebra $\frak{p}_2$ preserves the \Mink-light
cone, then with the translations $\P_j$ and others span the whole
algebra $\frak{im}(1,3)$. Moreover, by combining the Galilei and
Carroll boosts $\K^\frak{g/c}_i$, it follows the Poincar\'e algebra
$\frak{p}$, which is related to the \dS\ algebra $\frak{d}_+$  by
the combination of its translations and pseudo-translations of
$\frak{p}_2$.

The sign of the pseudo-translations $H'$, $\P'_i$ and the Carroll
boosts are determined by the algebraic relations. Since the correct
dimensions must
 be kept for all
generators and algebraic relations, once the sign of a generator is
changed, the sign(s) of relevant generators must also be changed so
that the original algebraic relations are still there as time
reversal or space inversion are made.

In this paper, the order of the time and space coordinates are
fixed. If it is changed, similar algebraic relations must occur. In
some cases this  makes sense. For instance, the kinematic algebras
$\frak{so}(2,3)$ with different orders can be regarded as the
conformal extensions of relativistic algebras on the space of one
dimension lower and so on \cite{C3}.

The combinatory approach is very different from the contraction
approach \cite{IW, Bacry}. Although we have mainly focused on the
Lie algebraic aspects of kinematics, the results already indicate
that this is also the case for corresponding groups, geometries
with metrics and global aspects.



\begin{acknowledgments} The authors would like to thank
Profs/Drs  Z. Chang, Q.K. Lu, Z.Q. Ma, J.Z. Pan, X.A. Ren, X. C.
Song, Y. Tian, S.K. Wang, K. Wu, X.N. Wu, Z. Xu, X. Zhang and C.J.
Zhu for valuable discussions. This work is partly supported by
NSFC (under Grants Nos. 10701081, 10775140, 10505004, 10675019),
NKBRPC(2004CB318000),  Knowledge Innovation Funds of CAS
(KJCX3-SYW-S03), and Beijing Jiao-Wei Key project
(KZ200810028013).
\end{acknowledgments}

\end{document}